\documentclass[twocolumn,floats,prb,aps,showpacs,10pt]{revtex4-1}
\usepackage[final]{graphicx}
\usepackage[dvipsnames]{color}
\usepackage{dcolumn}
\usepackage{hhline}
\DeclareGraphicsExtensions{.pdf,.eps}

\begin{document}

\title{First-principles GW calculations for fullerenes, porphyrins, phtalocyanine,
 and other molecules of interest for organic photovoltaic applications}

\author{X. Blase, C. Attaccalite, V. Olevano}

\affiliation{ Institut N\'{e}el, CNRS and Universit\'{e} Joseph Fourier,
B.P. 166, 38042 Grenoble Cedex 09, France, \\ and \\
European Theoretical Spectroscopy Facility (ETSF), Grenoble, France.} 

\date{\today}

\begin{abstract}
We evaluate the performances of \textit{ab initio} GW calculations for the 
ionization energies and HOMO-LUMO gaps of thirteen gas phase molecules of 
interest for organic electronic and photovoltaic applications, including 
the C$_{60}$ fullerene,  pentacene, free-base porphyrins and phtalocyanine,
PTCDA, and  standard monomers such as thiophene, fluorene, benzothiazole or 
thiadiazole.  Standard G$_0$W$_0$ calculations, that is starting from eigenstates 
obtained with local or semilocal functionals, significantly improve the ionization
energy and band gap as compared to density functional theory Kohn-Sham results, but 
the calculated quasiparticle values remain too small as a result of overscreening. 
Starting from Hartree-Fock-like eigenvalues provides much better results and is 
equivalent to performing self-consistency on the eigenvalues, with a resulting 
accuracy of 2-4$\%$ as compared to experiment. Our 
calculations are based on an efficient gaussian-basis implementation of GW 
with explicit treatment of the dynamical screening through contour deformation 
techniques.
\end{abstract}



\maketitle

\section{Introduction}

The flexibility in the synthesis of novel molecules and polymers is an important 
advantage of organic photovoltaics as compared to the inorganic  route
\cite{Mayer07,Sariciftci03}. Despite a rather limited quantum efficiency, 
the possibility to tailor the solubility, cristallinity and  electronic properties 
of the building molecular units is offering much means to improve on the actual best cells,
such as those based on the combination of acceptor fullerene derivatives 
and derivatives of polythiophene as donors \cite{PCBMP3HT,Brabec10}. 
In particular, it has been shown that there are strong correlations
between the ``band offsets" at the donor/acceptor interface and the open circuit voltage 
or the driving force for separating the hole and electron of the photoinduced  
excitons \cite{OCV,Kooistra07}.  
The ability to tune the electronic affinity and ionization energy of the donor 
and acceptor molecules, under the constraint that sun light absorption should be
kept as large as possible, is a current and intense field of research 
\cite{Schueppel08,Lincker08,Park09,Chen09}.
There is therefore much interest in developing efficient quantum simulation methods allowing
to provide the spectroscopic and optical properties of standard molecules with both a reasonnable
computer cost and accuracy.

For isolated molecules, an excellent trade-off between computer cost and accuracy for the
calculations of the ionization energy and electronic affinity can be found with the so-called 
$\Delta$SCF approach using hybrid functionals such as PBE0 and B3LYP obtained by admixture
of a fixed amount of  Fock exchange  \cite{Becke93,Perdew96}.  However, these techniques cannot
be used for extended systems such as bulk semiconductors, molecules deposited on a surface
or in solution, and the percentage of Fock exchange needed for obtaining good results 
with these functionals is expected to change from isolated molecules to bulk systems. 
For the same reasons, the ``Kohn-Sham"  ionization energies, electronic affinities and 
band gaps as obtained from the eigenvalues of the Hamiltonian may be certainly improved
with hybrid functionals as compared to (semi)local ones, but again the amount of Fock exchange
needed to get accurate results may change from one system to another.

A technique based on many-body perturbation theory (MBPT), namely the GW approximation 
\cite{Hedin65,Hybertsen86,Godby88,Onida02}, has shown excellent results for the evaluation 
of the band edges and band gaps of extended bulk systems \cite{Aulbur}. 
Distinct from the perturbative techniques
developed by the quantum chemistry community to build up correlations from the
Hartree-Fock solution \cite{zabo}, such an approach is generally derived from functional 
derivative techniques \cite{Schwinger59,Hedin65} yielding an exact (non-perturbative) set 
of self-consistent (closed) relations between the one-body Green's function G, the
polarizability $P$, the dynamically screened Coulomb  potential W, the
``exchange and correlation" self-energy $\Sigma$ and the so-called vertex
corrections $\Gamma$, which is related to the variation of 
the self-energy with respect to an external pertubation.
In practice, neglect of vertex corrections leads to the so-called ``GW" approximation
for the self-energy
which can be loosely described as a generalization of the Hartree-Fock
method by replacing the bare Coulombian
with a dynamically screened Coulomb interaction.  
The ingredients needed to proceed through the GW calculations pave further the way to 
Bethe-Salpeter calculations \cite{Onida02} aiming at exploring optical absorption 
spectra as an alternative to time-dependent DFT. While decades of expertise exist
for appraising the performances of the GW approximation in the case of extended
bulk systems, the application of such MBPT approaches to organic molecules in the
gas phase, and in particular molecules of interest for photovoltaic applications 
\cite{Dori06,Tiago08,Palumno09,Umari09,Ma09}, remain extremely scarce, a situation that can be mostly
attributed to the associated computational cost for molecules 
such as fullerene derivatives or porphyrins containing several dozens of atoms. 
As a result, an understanding of the merits of such an approach in the case of 
organic molecular systems, as compared to well-established quantum chemistry 
techniques, is still in its infancy.

\begin{figure*}
\begin{center}
\includegraphics*[width=\textwidth]{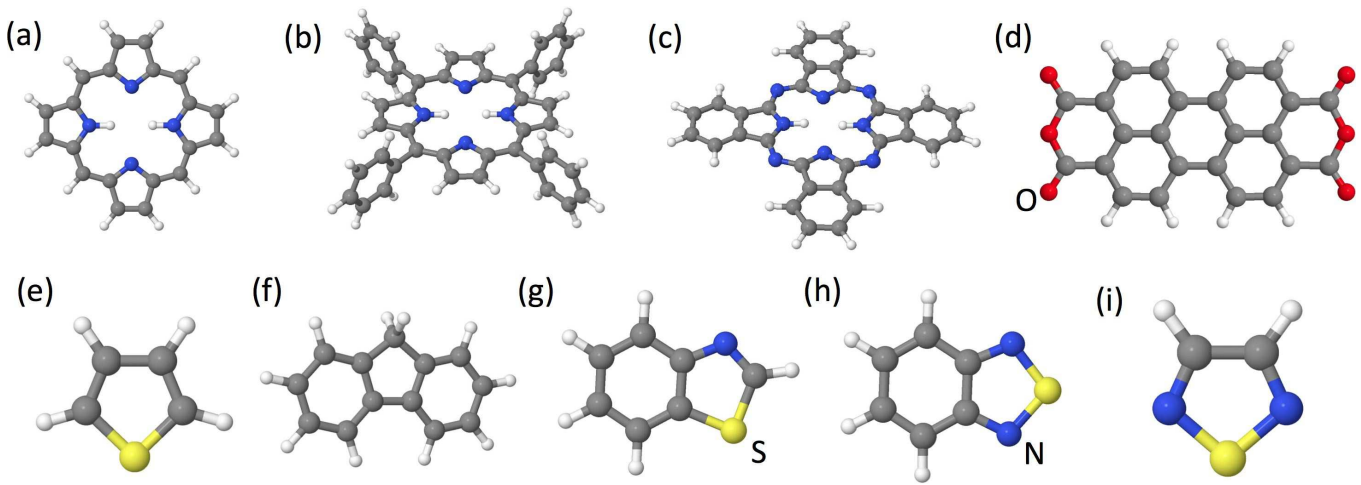}
\caption{(Color online) Symbolic representation of (a)   21H,23H-porphine (H$_2$P), (b) 
tetraphenylporphyrin (H$_2$TPP), (c) phtalocyanine (H$_2$Pc) (d) 3,4,9,10-perylene tetracarboxylic 
acid dianydride (PTCDA), (e) thiophene, (f) fluorene, (g) benzothiazole, (h) 2,1,3-benzothiadiazole
and (i) 1,2,5-thiadiazole. Small white atoms are hydrogen atoms, grey atoms
are carbon atoms while red/blue/yellow atoms are oxygen/nitrogen/sulfur atoms respectively. }
\label{fig1}
\end{center}
\end{figure*}

We present in this work a GW study of the quasiparticle properties of thirteen of the most
standard molecules involved in organic electronic and photovoltaic devices, including 
the $C_{60}$ fullerene, the  free-base 21H,23H-porphine (H$_2$P), tetraphenylporphyrin 
(H$_2$TPP),  and phtalocyanine (H$_2$Pc), and the 3,4,9,10-perylene tetracarboxylic acid
dianydride (PTCDA) (see Fig.~\ref{fig1}).  We also study anthracene, tetracene, and pentacene, 
$\pi$-conjugated molecules of interest for organic electronics, even though not as such 
for optical applications, and for which experimental band gap data are available. 
Finally, the tiophene, fluorene, benzothiazole,  2,1,3-benzothiadiazole and 1,2,5-thiadiazole
 monomers, 
building blocks of common donor polymers, are also investigated \cite{DAD,choice}. 
Our results suggest that while the standard  non-self-consistent G$_0$W$_0$ calculations
based on Kohn-Sham eigenstates with (semi)local functionals certainly improves on the DFT
results, the G$_0$W$_0$ ionization energy and  HOMO-LUMO gap remain underestimated as
compared to experiment.  A simple partial self-consistency on the eigenvalues only,
or the use of Hartree-Fock-like eigenvalues in a one-shot G$_0$W$_0$ calculation, 
allows to obtain much improved results. We show in particular that these simple schemes 
lead to an average error of $\sim$0.3 eV for the ionization energies and 0.1-0.2 eV for 
the band gaps.

Our paper is organized as follows. In section (II), we briefly describe our implementation of the
GW formalism within a gaussian-basis, including details about the evaluation of the Coulomb matrix 
elements.  In section (III), our results for the ionization energy and HOMO-LUMO gap of 
selected molecules are presented and compared to existing  experimental results. 
The importance of a simple self-consistency on the eigenvalues is discussed. 
Section (IV) describes a simplified non-self-consistent approach based on an 
approximate perturbative 
Hartree-Fock starting point for building the Green's function and screened
Coulomb potential. We conclude in section (V).

\section{Methodology}

Our code is based on a gaussian-basis implementation of the GW formalism and builds on a
previous implementation of calculating the inverse dielectric matrix using numerical strictly
localized orbitals \cite{fiesta}. To avoid dealing with numerical basis, the present 
implementation now expands the needed two-point operators (bare and screened Coulomb potentials, 
susceptibilities, etc.) on an ``auxiliary" gaussian basis composed of one-center atomic-like 
orbitals, with real spherical harmonics for the  angular part and a radial dependence composed 
of gaussian functions. 
The use of such an auxiliary basis, commonly implemented in several DFT quantum chemistry
codes to express the charge density for ground-state or excited-state \cite{Bauernschmitt97}
calculations, allows to greatly speed up the evaluation of e.g. the Coulomb matrix elements. 
We discuss these points in the following subsections.

\subsection{General formalism}

With the notations of Ref.~\onlinecite{Rohlfing95},  we introduce for any two-point function
$f(\textbf{r,r'})$ the $<f>$ and $[f]$ matrices in the auxiliary basis related through:

\begin{eqnarray*}
 \lbrack f \rbrack_{\mu,\nu}  &=& \int \!\! \int d{\bf r}\, d{\bf r'}
\; \mu^*({\bf r}) f({\bf r,r'}) \nu({\bf r'})  \\
 <f> &=&  S^{-1} \lbrack f \rbrack S^{-1}  \\
 f({\bf r,r'}) &=& \sum_{\mu,\nu} \mu({\bf r}) <f>_{\mu,\nu} \nu^*({\bf r'})
\end{eqnarray*}

\noindent where $\mu$ and $\nu$ are elements of the basis and $S$ is the overlap matrix.
The standard Dyson equation relating the dynamically screened Coulomb
potential $W(\omega)$ to the bare Coulomb one ($v$)  can then be written:

\begin{eqnarray*}
<W(\omega)> & = & <v> + <v> \lbrack \chi^0(\omega) \rbrack <W(\omega)>
\label{dyson}
\end{eqnarray*}

\noindent  with $\chi^0$ the unscreened (free-electron) susceptibility:

\begin{eqnarray*}
\lbrack \chi^0(\omega) \rbrack_{\mu,\nu} & = &  \sum_{spins} \sum_i^{occ}
\sum_{j}^{unocc}  
 <\phi_i | \mu | \phi_j> < \phi_j | \nu | \phi_i>   \\
 & \times & \left(  {1 \over \omega  +  \varepsilon_i-\varepsilon_j + i\delta   }
    - { 1 \over \omega  -  \varepsilon_i+\varepsilon_j - i\delta} \right) 
\label{chi0bis}
\end{eqnarray*}

\noindent where $\delta=0^+$.
The input $(\phi_i,\varepsilon_i)$ are one-body eigenstates and related eigenvalues
traditionaly taken as the Kohn-Sham solutions of a ground-state DFT calculation. In
the present paper, we start with a standard  DFT/LDA calculation but as discussed below,
this may not constitute the best starting point for molecular systems.
The knowledge of the dynamical screened Coulomb potential $W(\omega)$ allows to build
the non-local and energy dependent self-energy operator $\Sigma$, which accounts for 
exchange and correlation in the present quasiparticle formalism \cite{Hedin65} and reads: 

\begin{eqnarray*}
 \Sigma^{GW}(\bf{r,r'}|E) &=& { i \over 2\pi } \int d\omega e^{i 0^+ \omega}
    G({\bf{r,r'}}|E+\omega) { W}({\bf{r,r'}}|\omega) \\
 G({\bf{r,r'}}|\omega) &=& \sum_n \phi_n({\bf{r}}) \phi_n^*({\bf r'}) / (\omega - {\varepsilon}_n \pm i\delta)
\label{sigma}
\end{eqnarray*}

\noindent where the time-ordered Green's function $G$ is again built from the $(\phi_i,\varepsilon_i)$ 
eigenstates. The sign of the $\delta$ infinitesimal insures that the occupied (unoccupied) 
states correspond to poles in the fourth (second) quadrants. Again, the choice of the
``best" input $(\phi_i,\varepsilon_i)$ for the building of G  will be discussed below.

This implementation is formally equivalent to that of Ref.~\onlinecite{Rohlfing95}
except that we go beyond the plasmon-pole model and proceed with the explicit 
calculation of the frequency integral for the correlation part of the self-energy, 
$\Sigma_c^{GW} = \Sigma^{GW} - \Sigma_x$, with $\Sigma_x$ the Fock operator. We use contour deformation 
techniques with an integration along the imaginary axis complemented by the evaluation of the poles
in the first and third quadrant for states away from the band edges \cite{Godby88,Farid99}:

\begin{eqnarray*}
 \Sigma_c^{GW}({\bf{r,r'}}|E) & = &   \sum_n \phi_n({\bf{r}}) \phi_n^*({\bf{r'}})
     \mathcal{V}_n({\bf{r,r'}}|E)
\end{eqnarray*}

\noindent with, introducing ${\tilde W} = W -v$, $E_F$ the Fermi level, and 
${\theta}$ the Heaviside step function:

\begin{eqnarray*}
 \mathcal{V}_n({\bf{r,r'}}|E) &=& {\tilde W}({\bf{r,r'}}| \varepsilon_n - E)
    \left[ {\theta}(E-\varepsilon_n) - {\theta}(E_F-\varepsilon_n) \right]  \\
 && -  \int_0^{+\infty} { d\omega \over \pi }
     { E -\varepsilon_n \over (E -\varepsilon_n)^2 + \omega^2 }
  {\tilde W}({\bf{r,r'}}| i\omega)
\label{contour}
\end{eqnarray*}

\noindent
A change of variable allows to fold the smooth function ${\tilde W}(i\omega)$ onto
the finite $[0,1]$ interval where Gaussian quadrature with as little as 12
gaussian points is sufficient to reach convergency. An analytically integrable tail
is added/subtracted to avoid instabilities with the integrand for $\omega \rightarrow 0$
when $E \rightarrow \varepsilon_n$. 

The first order perturbation theory self-energy correction to the DFT Kohn-Sham eigenvalues is extrapolated 
to the quasiparticle energies by  a Taylor expansion, namely: 

$$
 \varepsilon^{QP}_n = \varepsilon_n + Z_n <\phi_n | 
    \Sigma^{GW}(\varepsilon_n)-V_{xc}^{LDA}| \phi_n >
$$

\noindent where $Z_n$ is the renormalization factor defined as: 

$$
 1/Z_n = 1 - \left[ \partial \Sigma^{GW} / \partial E \right]_{\varepsilon=\varepsilon_n}.
$$

\noindent with $(\varepsilon_n,\phi_n)$ the LDA Kohn-Sham eigenvalues and eigenstates 
in the present case.

\subsection{Gaussian basis}

The auxiliary basis used to expand  the two-point functions reads: 
$\; \mu({\bf r})$ = exp$(-\alpha r^2) r^l R_l^m({\hat r})$ with
$R_l^m({\hat r})$ the real-spherical harmonics and $({\hat r})$ the angular
components of the {\bf r}-vector. It is computationally more efficient
to work with the  $R_l^m$ instead of the standard $Y_l^m$ complex harmonics 
with the following relation:

$$
 R_l^m({\hat r}) = \left\{ 
   \begin{array}{ll}
      \left[ Y_l^m({\hat r}) + (-1)^m Y_l^{-m}({\hat r}) \right] / \sqrt{2} & \mbox{(m $>$ 0)} \\
      Y_l^m({\hat r})  & \mbox{(m = 0)} \\
      \left[ Y_l^{-m}({\hat r}) - (-1)^m Y_l^m({\hat r}) \right] / \sqrt{2} & \mbox{(m $<$ 0)} 
   \end{array} 
  \right.
$$

\noindent
The products $r^l R_l^m({\hat r})$ yield the standard expressions (\textit{x,y,z,xy,yz,
x$^2$-y$^2$}, etc.) for the \textit{p,d}, etc. orbitals (within constant factors).  
We briefly recall that the main advantage of a gaussian radial part (as compared
to numerical or Slater-type orbitals) is that the product of two gaussians centered on atoms
$A$ and $B$ with decay coefficients $\alpha_1$ and $\alpha_2$ yields a gaussian 
centered on $C= (\alpha_1 A + \alpha_2 B)/(\alpha1+\alpha_2)$  with a decay constant
$\gamma = \alpha_1 \alpha_2 /(\alpha_1+\alpha_2)$. Further, the $r^l R_l^m({\hat r})$ 
can easily be ``shifted" from one center to another with for sake of illustration:

\begin{eqnarray*}
 (x-x_A)(y-y_A) &=& (x-x_C)(y-y_C)  \\
                &+&   (y_C - y_A) (x - x_C) \\
                &+& (x_C - x_A) ( y-y_C) +  \text{constant}, 
\end{eqnarray*}

\noindent showing that a  \textit{d}$_{xy}$ orbital centered on $A$ can be easily expressed
as a function of (\textit{s,p}) and \textit{d}$_{xy}$ orbitals centered on $C$. Such 
trivial expressions allow to express multi-center overlaps in terms of one-center 
integrals. 

In the present work,
our calculations start with a DFT calculation of the structural and electronic 
properties of the molecules of interest using the Siesta package \cite{siesta}.  We use a 
``double-$\zeta$+polarization" (DZP) basis \cite{TZDP} and standard norm-conserving
pseudopotentials. Since the  Siesta package uses ``numerical" orbitals, we first fit
the numerical radial part by up to five contracted gaussians \cite{localization} in
order to exploit the relations briefly sketched above. As such, both the ``ground-state"
DFT basis and the auxiliary basis are based on gaussians.  
Beyond the analycity of the gaussian basis, our choice was also motivated by the possibility 
of using eigenstates generated by standard chemistry codes with all electron approaches
and/or hybrid functionals, providing for some systems possibly a better starting point for 
MBPT calculations (see discussion below).  We labeled our code ``Fiesta" as an attempt to 
extend the ``Siesta" package to excited state properties.

Contrary to the planewave case,  the auxiliary basis for the two-point response functions is larger 
than the ground-state basis.  Following  Kaczmarski and coworkers \cite{Kaczmarski10}, we typically adopt 
for first raw elements such as carbon, nitrogen and oxygen, 4 \textit{s,p,d} sets of gaussian 
orbitals, that is 36 orbitals per atom, while 3 \textit{s,p,d} gaussian orbitals are sufficient 
for hydrogen. We show below that such a basis is large enough for the studied organic systems.
In the case of sulfur, \textit{f}-channel orbitals are added.
With such a basis, a typical G$_0$W$_0$ calculation with full dynamics for our largest molecule
(H$_2$TPP) can be performed within one day on a single standard processor. Better timings and scaling
may be obtained upon implementing the recently introduced techniques allowing to avoid summation
over the conduction states \cite{Bruneval08,Umari09,Giustino10,Berger10}, or techniques decoupling
the sum over valence and conduction states \cite{Foerster}, even though the number of unoccupied 
states is rather limited with standard DZP or larger TZDP basis.

The choice of the ``optimal" $\alpha$-coefficients, controlling the localization of the basis
orbitals, is a difficult question. 
Auxiliary basis have been implemented in many quantum chemistry codes in order to fit the charge
density and speed up the calculation of the Coulomb integrals. The coefficients of
the charge density on the auxiliary basis are optimized using "identity rules" 
\cite{identityrules} but not the decay coefficients in the exponentials. Years of expertise in the
quantum chemistry community yielded reliable auxiliary basis for the periodic table and numerous 
tests have shown that high precision can be obtained with such basis provided that they be
sufficiently large. 

Since the auxiliary basis must project onto products of Kohn-Sham orbitals, optimized basis
for all-electron calculations cannot be straightforwardly used for GW calculations starting
from ground-state calculations with pseudopotentials. The same guiding lines can however be
followed. We adopt in particular the idea
of a ``tempered" basis \cite{Reeves63,Ruedenberg73,Cherkes09} suggesting that it is better to
generate a chain of $\alpha$ parameters such that:  $\; \alpha_{i+1} / \alpha_{i}$ = constant,
rather than spreading them uniformly between $\alpha_{min}$ and $\alpha_{max}$. 
Such a scheme hinges on the facts that the overlap of two gaussian orbitals is a function of
their alpha coefficient ratio and that maintaining a constant overlap between ``adjacent"
gaussians allows to better span the associated Hilbert space \cite{Cherkes09}. As such, the $\alpha_{min}$,
$\alpha_{max}$ and number of gaussian per l-channel being chosen, the other gaussian coefficients
are automatically generated.

We adopt the basis proposed by Kaczmarski and coworkers \cite{Kaczmarski10}, that is namely
gaussians with localization parameters of (0.2,0.5,1.25,3.2) a.u. for the (\textit{s,p,d}) channels
of C, O, and N atoms, and gaussians with $\alpha$=(0.1,0.4,1.5) a.u. for hydrogen. As shown in Table I 
in the case of anthracene, H$_2$P porphyrin and C$_{60}$, changing the $\alpha_{min}$ and $\alpha_{max}$ 
values, or increasing the number of gaussians in the basis, does not change significantly the
results. The case of $C_{60}$ shows however that reducing the number of gaussians to 3 per l-channel
yields a significant error on the band gap.  We will show below that the results 
obtained with the present implementation compares rather  well with previous available calculations 
based on another gaussian basis, planewaves (PWs) or combination of gaussians, PWs and Wannier functions. 

\begin{table}
\begin{tabular}{|l|c|c|c|c|c|c|}
   \hline
 auxiliary  basis & \multicolumn{2}{c|}{anthracene} & \multicolumn{2}{c|}{H$_2$P} & \multicolumn{2}{c|}{C$_{60}$} \\
   \hline
ng in $\alpha_{min}$  $\rightarrow$  $\alpha_{max}$ & IE & gap & IE & gap & IE & gap \\
   \hline
3 in 0.2 $\rightarrow$ 3.2   & 6.83   & 6.02   & 6.49   & 4.67  & 7.21  & 4.08  \\
   \hline
4 in 0.2 $\rightarrow$ 3.2   & 6.89   & 6.15   & 6.56   & 4.79  & 7.29   & 4.44  \\
   \hline
5 in 0.2 $\rightarrow$ 3.2   & 6.86   & 6.14   & 6.52   & 4.76  & 7.30  & 4.37  \\
   \hline
4 in 0.15 $\rightarrow$ 3.2   &  6.89  & 6.15   &  6.52  & 4.74  &  7.40  & 4.47      \\
   \hline
5 in 0.15 $\rightarrow$ 3.2   & 6.82   & 6.06   &  6.56  & 4.77  & 7.29  & 4.36  \\
   \hline
5 in 0.15 $\rightarrow$ 3.5   & 6.83   & 6.08   &  6.51  & 4.75  & 7.28  & 4.33  \\
   \hline
\end{tabular}
\caption{Evolution of the ionization (IE) and band gap  energies of selected molecules 
as a function of the carbon auxiliary basis, changing the number (ng) of gaussians per l-channel, 
the $\alpha_{min}$ and $\alpha_{max}$ coefficients. Results are in eVs.}
\label{tablebasis}
\end{table}

We conclude this section related to the auxiliary basis by mentioning an important numerical
aspect related to the overcompletness of the generated non-orthogonal gaussian basis. While 
the basis on a given atom can be easily orthogonalized using e.g. a Gram-Schmidt procedure,
the overlap between the most diffuse orbitals on adjacent atoms tend to be also rather large
yielding an overlap S matrix ``nearly singular".  Following the strategy developed 
in the case of product-basis \cite{Aryazetiawan,Foerster}, we rotate our basis to 
that of the eigenvectors of the overlap S-matrix from which we remove the eigenvectors
with eigenvalue smaller than typically $10^{-5}$. In the present case of auxiliary basis, such
a truncation does not reduce significantly the size of the basis, but avoid the potential 
numerical instability associated with inverting the nearly-singular S-matrix and the amplification of 
errors associated with the $\; <v> = S^{-1} [v] S^{-1}$ transformation (see above). 
The cost of rotating the Coulomb  and $< \phi_i | \beta | \phi_j>$ matrix elements from
the original one-center auxiliary basis ($\beta$) to the (filtered) S-eigenvectors basis
scales as $N^3$ and represents a marginal part of the CPU time.

\subsection{Coulomb matrix elements}

An important ingredient is the evaluation of the Coulomb matrix elements between
two auxiliary basis orbitals localized on two different atoms. Exploiting the
properties of the Fourier transform (FT) of gaussian-based orbitals, namely:

\begin{equation}
FT \left[ e^{-\alpha r^2} r^l R_l^m ({\hat r}) \right] = 
    C e^{-\gamma q^2} q^l R_l^m ({\hat q})  
\end{equation}

\noindent with $\gamma=1/4\alpha$ and $({\hat r},{\hat q})$ the angular components of
the ({\bf r,q})-vectors in direct and reciprocal space respectively (C is a constant),
the Coulomb matrix elements reduce to a sum of terms built
from the product of one-center overlaps of three real-spherical harmonics
$<{R_l^m}{R_{l'}^{m'}}|R_L^M>$ (related to Gaunt coefficients with $|l-l'| \le  L \le (l+l')$)
times radial integrals I(l,l';L) of the form:

$$
I(l,l';L) =  \int_0^{\infty} dq \; e^{-\zeta q^2} q^{\mu} J_{\nu}(-\beta q^2) 
$$

\noindent The  $<{R_l^m}{R_{l'}^{m'}}|R_L^M>$ factors  are pretabulated.
The oscillatory behavior of the Bessel function of the first kind $J_{\nu}$ makes
the direct numerical evaluation rather unstable. We prefer to notice that $I (l,l';L)$
is straighforwardly related to the $_1F_1$  confluent hypergeometric functions 
\cite{Gradsteyn}  which, for the needed $(l,l')$ values, can be expressed in terms of 
simple functions such as the error function (erf) or the Dawson integral:    
 F(z) = $\sqrt{\pi}$ exp(-z$^2$) erfi(z)/2, with  erfi(z) = erf(iz)/i,   
for which rapidly convergent serial expressions exist
\cite{Faddeeva}. This is an important advantage of the auxiliary basis  approach that 
the evaluation of the off-site Coulomb matrix elements is not a costly part of the
present GW implementation.

\begin{table*}
\begin{tabular}{||p{0.15\textwidth}|p{0.15\textwidth}|p{0.15\textwidth}|p{0.15\textwidth}|p{0.15\textwidth}|p{0.15\textwidth}||}
\hhline{|t:======:t|}
\multicolumn{6}{||c||}{\textbf{Ionization energy}} \\
\hhline{||------||}
           &  LDA-KS &  G$_0$W$_0$(LDA) & GW     & G$_0$W$_0$(HF$_{\rm diag}$) & Experiment \\
\hhline{||------||}
anthracene &   5.47   &  6.89  & 7.06   &  7.03   &  7.4$^a$  \\
tetracene &  5.15    &  6.37  & 6.51   &  6.48   &  6.97$^a$  \\
pentacene  &   4.94   &  5.98  &  6.12  &  6.08   &  6.6$^a$  \\
C$_{60}$   &   6.37   &  7.28  &  7.41  &  7.41   &  7.6$^a$    \\
PTCDA      &   6.65   &  7.57  &  7.68   &  7.67  &  8.2$^b$  \\
H$_2$P        &   5.64   &  6.55  &  6.70    & 6.72  &  6.9$^a$  \\
H$_2$TPP      &   5.40   &  6.09  &  6.20   & 6.24    &  6.4$^a$  \\
H$_2$Pc       &   5.56   &  6.08  & 6.10   &  5.93   &   6.4$^c$\\
thiophene  &   6.15   &  8.37     &  8.63   &  8.64   &  8.8$^a$  \\
fluorene   &   5.92  &  7.44     &  7.64   &  7.64   &  7.9$^a$  \\
benzothiazole &  6.33 &   8.20    &  8.48   &  8.50   &  8.8$^a$  \\
thiadiazole       &  7.22   & 9.65 & 9.89 & 9.90   & 10.1$^d$ \\
benzothiadiazole  & 6.55 &  8.31  &  8.56   &  8.57   &  9.0$^a$  \\
\hhline{||------||}
MAE        &   1.83 (23$\%$) &  0.47  (6$\%$)   &  0.30 (3.8$\%$)  &  0.31 (4.0$\%$)  &        \\
\hhline{|b:======:b|}
\end{tabular}
\caption{Ionization energies in eV as obtained from the Kohn-Sham eigenvalues (LDA-KS), 
from non-self-consistent G$_0$W$_0$(LDA) calculations, from a GW calculation with self-consistency
on the eigenvalues (GW), and from a non-self-consistent G$_0$W$_0$ calculation starting from 
Hartree-Fock-like eigenvalues (G$_0$W$_0$(HF$_{\rm diag}$), see text). 
MAE is the average mean error in eV. The average error in percent as compared to the experiment
is indicated in parenthesis.
\noindent $^a$Ref.~\onlinecite{NIST}. 
\noindent $^b$Ref.~\onlinecite{Dori06}.
\noindent $^c$Ref.~\onlinecite{H2Pc_exp}. 
\noindent $^d$Ref.~\onlinecite{thiadiazole_exp}.
}
\label{tableIE}
\end{table*}

\section{Results}

\subsection{Ionization energies}

\begin{figure}
\begin{center}
\includegraphics*[width=0.45\textwidth]{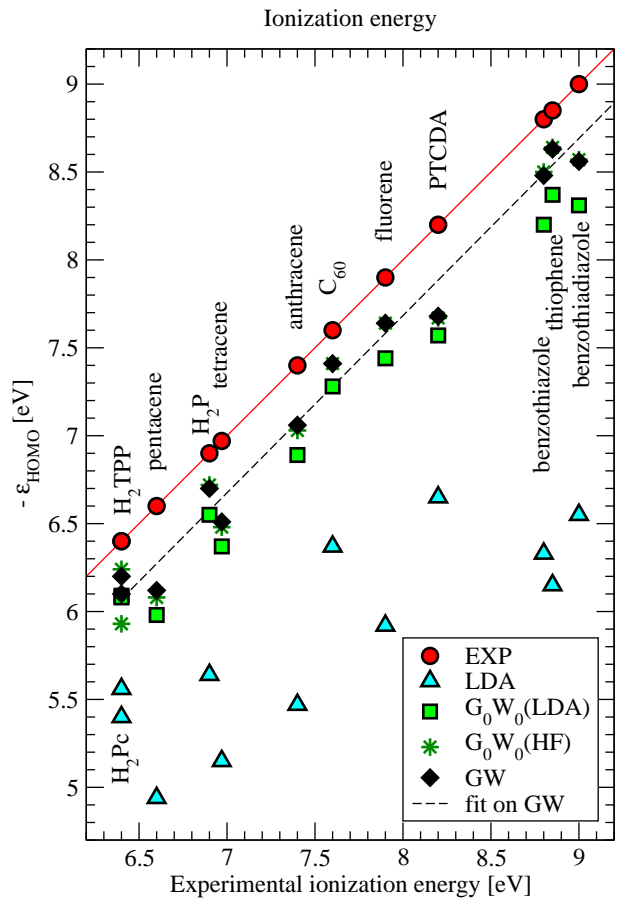}
\caption{(Color online) Experimental and theoretical ionization energies in electronvolts.
Red circles: experimental values; light blue triangles up: LDA Kohn-Sham HOMO energy; 
green squares: non-self-consistent G$_0$W$_0$(LDA) value; black diamonds:
GW value with self-consistency on the eigenvalues; green stars: non self-consistent
G$_0$W$_0$(HF$_{diag}$) (see text). The black dashed line is a least-square fit of the
GW results. The figure has been formatted so as to preserve 
the same physical scale on both axis.}
\label{fig2}
\end{center}
\end{figure}


We start by exploring the ionization energy of our selected molecules. While
experimental data for the electronic affinity of molecules in the gas phase are 
scarce, accurate measured ionization energies are much more common \cite{NIST}. 
Experimental
ionization energies are represented by red circles in Fig.~\ref{fig2} and are given in the
last column of Table II.  The DFT-LDA ionization energies, as obtained from the 
opposite sign of the Kohn-Sham highest occupied (HOMO) energy level, are clearly much 
too small, with an average error of 1.83 eV or 23$\%$ (see blue triangles in Fig.~\ref{fig2}). 
Very similar results are obtained using the HOMO energy value as obtained with a 
semilocal functional such as PBE \cite{PBE}. 

We now turn to G$_0$W$_0$(LDA) calculations, that is non-self-consistent calculations
with the Green's function and screened Coulomb potential directly built from
the LDA Kohn-Sham eigenstates and eigenvalues. The analysis of the results
(column 3 Table I and green squares in Fig.~\ref{fig2}) shows that the ionization 
energies are greatly improved, with an average error of 0.47 eV, that is a much reduced 
6$\%$ error. 

Even though in much better agreement with experiment than LDA or PBE, the discrepancies are
still sizeable.  As shown below, part of the problem 
originates in that the ``starting" LDA HOMO-LUMO gap is dramatically too small for 
isolated molecules, inducing a large overscreening. To avoid using some arbitrary scissor
operator to open the HOMO-LUMO gap in calculating the susceptibility, we rather perform a 
restricted self-consistency by reinjecting the corrected eigenvalues in G and W up to
convergency.  As a matter of fact, no more 
than three or four iterations are needed to reach convergency within 0.01 eV.
Such an approximation is labeled GW in the following.  This is not a full 
self-consistent approach as the eigenstates are not updated, with the advantage that the
computational cost keeps reasonnable. Full self-consistency without vertex corrections 
is still debated  and seems to yield for small molecular systems results that are not
as good as G$_0$W$_0$ non self-consistent runs \cite{Rostgaard10}.

The analysis of the results (fourth column Table II and black diamonds in Fig.~\ref{fig2}) clearly 
shows that the self-consistency on the eigenvalues improves the results for the ionization energy, 
reducing the average error from 0.47 eV (6$\%$) to 0.30 eV (or 3.8$\%$). 
Such a discrepancy is still sizeable but much better 
than the one obtained from the LDA Kohn-Sham HOMO energy.  
An interesting observation is that the final GW ionization energies gather much 
closer to a straight line  (dotted black line on Fig.~\ref{fig2}) parallel to the first diagonal 
(red ``experimental" line) than the LDA data which are much more spread. On a pragmatical
point of view, this means that the band offset between two molecules will strongly benefit
from cancellation of errors in GW as compared to LDA. In particular, the remaining error 
($\sim$ 0.2 eV) on the ionization energy for C$_{60}$, 
the most standard acceptor, is nearly identical to the error on the 
ionization energies of porphyrins and phtalocyanines, which are commonly used donors.
We now show that self-consistency, even though limited to updating the eigenvalues only,
has an even larger effect on the magnitude of the HOMO-LUMO gaps.


\subsection{HOMO-LUMO gaps}

\begin{figure}
\begin{center}
\includegraphics*[width=0.45\textwidth]{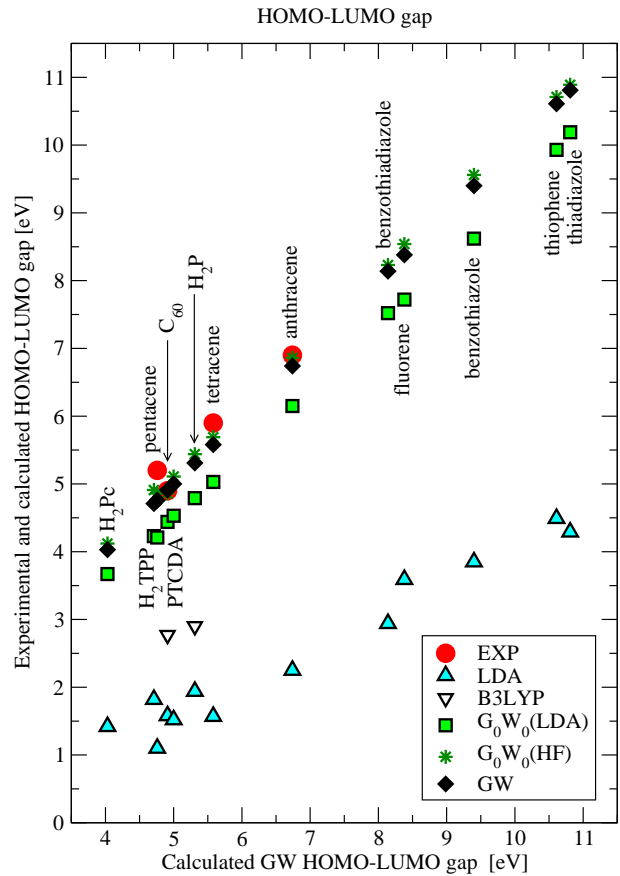}
\caption{(Color online) Experimental and theoretical HOMO-LUMO gaps in electronvolts.
Red circles: experimental values; light blue triangles up: LDA Kohn-Sham HOMO-LUMO gap; 
green squares: non-self-consistent G$_0$W$_0$(LDA) value; black diamonds:
GW value with self-consistency on the eigenvalues; green stars: non-self-consistent
G$_0$W$_0$(HF$_{diag}$) values (see text). The two down-pointing
empty triangles are B3LYP/6-31G(d) HOMO-LUMO gap values from 
Refs.~\onlinecite{b3lyp,ShucklaC60,ZhangC60,Nguyen} for $C_{60}$ and the H$_2$P porphin.   }
\label{fig3}
\end{center}
\end{figure}

Due to the lack of experimental values for the electronic affinity, experimental quasiparticle
HOMO-LUMO gaps (red circles in Fig.~\ref{fig3}) are scarce so that we plot our results as a function of our ``best"
calculated HOMO-LUMO gaps, namely the GW ones. 
In the case of C$_{60}$, anthracene, tetracene, and pentacene for which experimental data are available,
we observe as expected that the LDA HOMO-LUMO gap (blue triangles) is too small. This is well known in the case of bulk
semiconductors but here the discrepancy is much larger, with an average error  
of  $\sim$ 4.1 eV or $71\%$.

The G$_0$W$_0$(LDA) HOMO-LUMO gaps (green squares) significantly improves with respect to LDA.
Comparing to available G$_0$W$_0$(LDA) data for this class of aromatic molecules, our calculated 
6.15 eV HOMO-LUMO gap for anthracene compares  well with the 5.97 eV values of Niehaus and 
coworkers, despite the differences in basis and the treatment of dynamical effects  \cite{Niehaus05}. 
Our G$_0$W$_0$(LDA) 4.79 eV and 4.23 eV HOMO-LUMO gaps for the H$_2$P and H$_2$TPP 
free-base porphyrins respectively 
compare further well with the 5 eV and 4.39 eV planewave results of Palummo and coworkers 
\cite{Palumno09}. Similarly, our G$_0$W$_0$(LDA) 4.44 eV band gap for $C_{60}$ is in  good
agreement  with the
real-space grid formulation of Tiago and coworkers \cite{Tiago08} yielding a band gap of 4.36 eV.
Such comparisons certainly underline the reliability of the present gaussian-basis implementation.
Our 4.53  eV band gap for PTCDA is however smaller than the 4.9 eV band gap found with a previous 
planewave GW calculation \cite{Dori06,convergency,homom1}.

Overall, we remark a systematic underestimation of the G$_0$W$_0$(LDA) HOMO-LUMO gap
with respect to the experiment, with an average error
for our test molecules of $\sim$0.75 eV or $13\%$. 
This contrasts with the case of bulk systems for which the results of G$_0$W$_0$(LDA) are generally 
in much better agreement with experimental values. Such a behavior can be analyzed by noticing 
that building the polarizabilities and screened Coulomb potential with LDA eigenvalues, that is
in particular with dramatically too small HOMO-LUMO gaps, leads to a significant overscreening. 
This induces too large a correlation correction ``G(W-V$^C$)" to the Hartree-Fock HOMO-LUMO gap, 
that is too small a HOMO-LUMO gap.

Even though much better than the Kohn-Sham HOMO-LUMO gap obtained with e.g. the B3LYP 
functional \cite{b3lyp} (see empty down triangles in Fig.~3), it is desirable to improve 
the results.  Following the simple scheme introduced 
above, performing self-consistency on the eigenvalues in G and W, the GW HOMO-LUMO gap
is further increased to reach much better agreement with experiment. 
The MAE is now reduced to 0.22 eV or 3.8$\%$ for our four test molecules. 
In the case of $C_{60}$, which is the most standard acceptor in organic photovoltaic
cells, the excellent agreement with experiment for the band gap value is rather satisfactory.
It is interesting to note further that the MAE of 0.22 eV for HOMO-LUMO gaps is close to the 
0.3 eV MAE obtained for the ionization energies, suggesting that the electronic affinity is 
quite well reproduced on the average. 

\section{A simple non-self-consistent G$_0$W$_0$ approach based on Hartree-Fock-like eigenvalues.}

We conclude this study by exploring a simple non-self-consistent G$_0$W$_0$ scheme starting from
an ``ansatz" Hartree-Fock (HF) calculation obtained by removing the exchange-correlation contribution to
the LDA eigenvalues and adding the diagonal part of the exchange operator in the LDA basis,
namely: 

$$
\epsilon^{``HF"}_n = \epsilon^{LDA}_n + < \phi^{LDA}_n | \Sigma_x - V_{xc}^{LDA} | \phi^{LDA}_n>,
$$

\noindent
where $\Sigma_x$ and $V_{xc}^{LDA}$ are the Fock and (semi)local exchange-correlation operators.
We label this very simple scheme G$_0$W$_0$(HF$_{\rm diag}$).
This  approximation was tested by Hahn, Schmidt and Bechstedt \cite{Hahn05} in the 
case of three small molecules (silane, disilane, water), arguing as we do that the Kohn-Sham 
eigenvalues are too bad a starting point to evaluate the time-ordered Green's function 
and the screened potential.  Such an approach is also  a variation on the G$_0$W$_0$(HF) 
scheme recently introduced in Ref.~\onlinecite{Rostgaard10} 
which was shown to yield the best ionization energies for small molecules. With increasing size
and number of electrons, the part of correlations in the self-energy is expected to become more 
important and using Hartree-Fock eigenstates/eigenvalues as a starting point for the much larger
systems we study may, in principle,  not be better
than using (semi)local functionals for generating the starting eigenstates.  
This is what we now explore.

\begin{table*}
\begin{tabular}{||p{0.15\textwidth}|p{0.15\textwidth}|p{0.15\textwidth}|p{0.15\textwidth}|p{0.15\textwidth}|p{0.15\textwidth}||}
\hhline{|t:======:t|}
\multicolumn{6}{||c||}{\textbf{HOMO-LUMO gap}} \\
\hhline{||------||}
           &  LDA-KS &  G$_0$W$_0$(LDA) & GW     & G$_0$W$_0$(HF$_{\rm diag}$) & Experiment \\
\hhline{||------||}
anthracene &   2.25   &  6.15   &  6.74    &  6.86  &  $6.9^a$  \\
tetracene &  1.57    &  5.03   &  5.58    &  5.69  &  5.9$^a$  \\
pentacene  &   1.10   &  4.21   &  4.76    &  4.86  &  $5.2^a$  \\
C$_{60}$   &  1.58  & 4.44   & 4.91  &  5.08 &  $4.9^a$  \\
\hline
MAE       &   4.10 (71$\%$) &   0.76 (13$\%$)   &  0.22  (3.8$\%$)   &   0.10 (2$\%$) &      \\
\hline
PTCDA      &  1.52   & 4.53   &  5.0    &  5.11   &    \\
H$_2$P        &    1.94  & 4.79    & 5.31     &  5.44   &    \\
H$_2$TPP      &  1.82   &  4.23   &  4.71  & 4.91    &    \\
H$_2$Pc       &  1.42   &  3.67   & 4.03   &   4.12  &    \\
thiophene  &  4.49   &  9.93  &  10.61    &   10.71    &    \\
fluorene   &  3.59  &  7.72   &  8.38  &    8.54    &    \\
benzothiazole   &  3.85   & 8.62   & 9.40  &  9.56   &    \\
thiadiazole        &  4.29   & 10.19  & 10.81  & 10.89  &  \\
benzothiadiazole    & 2.94  &   7.52     &  8.14     &   8.23   &    \\
\hhline{|b:======:b|}
\end{tabular}
\caption{HOMO-LUMO gap in eV as obtained from the Kohn-Sham eigenvalues (LDA-KS), 
non-self-consistent G$_0$W$_0$(LDA) calculations, a GW calculation with self-consistency
on the eigenvalues (GW), and a non-self-consistent G$_0$W$_0$ calculation starting from 
Hartree-Fock-like eigenvalues (G$_0$W$_0$(HF$_{diag}$), see text). 
MAE is the average mean error in eV for the anthracene, tetracene, pentacene and C$_{60}$ cases
for which experimental band gap data are available. 
The average error in percent as compared to the experiment is indicated in parenthesis.
\noindent $^a$Ref.~\onlinecite{NIST}.} 
\label{table-homolumo}
\end{table*}

For sake of comparison, we have studied the two small carbon-based conjugated molecules $C_2H_2$ 
and $C_2H_4$ which were investigated by Rostgaard and coworkers within their full $G_0W_0(HF)$ scheme.
The present  G$_0$W$_0$(HF$_{\rm diag}$) treatment increases the ionization energy by 3.48 eV and 3.80 eV for 
$C_2H_4$ and $C_2H_2$ respectively as compared to the LDA values.  Such corrections compare well 
with the 3.61 eV and 3.90 eV values obtained within the full $G_0W_0(HF)$ scheme of Rostgaard and 
coworkers (as compared to DFT/PBE), 
emphasizing the reliability of the present simplified approximation. 

As compiled in Table II and III (column 5) and in Figs.~2 and 3 (green stars), 
we do find as well that a single shot G$_0$W$_0$(HF$_{diag}$) calculation provides results
which are in good agreement with the full GW calculations with self-consistency on the
eigenvalues. 
In particular, the G$_0$W$_0$(HF$_{diag}$) calculations yield much better results
than the G$_0$W$_0$(LDA) scheme. Such a conclusion agrees with that of Rostgaard and coworkers
concluding that for small isolated molecules, the full G$_0$W$_0$(HF) scheme 
actually outperforms a 
full self-consistent GW  calculation where both eigenstates 
and eigenvalues are updated \cite{Rostgaard10}.

Within the present G$_0$W$_0$(HF$_{diag}$) approach, the MAE on the ionization energies
as compared to experiment is 0.31 eV, in good agreement with the 0.4 eV result of 
Ref.~\onlinecite{Rostgaard10} for small molecules. 
Such an agreement indicates that the present G$_0$W$_0$(HF$_{diag}$) implementation captures 
most of the features of a full $G_0W_0(HF)$ approach, suggesting that LDA and HF
eigenfunctions may not too different for this set of molecules, 
a conclusion oftened discussed in the literature.  Further, the error on 
the band gap, averaged on the calculated values for anthracene, tetracene, pentacene and
C$_{60}$, for which precise experimental data are available, is found to be as small as 0.1 eV 
(2$\%$ error). Such values compare very well with accurate quantum chemistry calculations with
a scheme, the GW formalism, which can be applied both to finite size and extended systems, and
allows to obtain not only the band edges, or frontier orbitals, but also the full quasiparticle 
spectrum (see note~\onlinecite{homom1}). 

\section{Conclusions}

We have explored the performances of several GW approximations for the calculation of the 
ionization energy and HOMO-LUMO gap of thirteen ``large" molecules of interest for photovoltaic 
applications, including $C_{60}$, free-base porphyrins and phtalocyanine, 
PTCDA and standard donor monomers  such as thiophene.  
Our calculations are based on a gaussian-basis implementation with full
dynamical effects through contour deformation techniques.  
Due to the dramatic error on the HOMO-LUMO gaps obtained with  (semi)local functionals, 
we find that the standard non-selfconsistent G$_0$W$_0$ calculations based 
on input LDA eigenstates performs rather poorly, in particular in evaluating the HOMO-LUMO gaps. 
A simple self-consistency on the eigenvalues used to build G and W provides much better results. 
As an even simpler scheme, a non-self-consistent G$_0$W$_0$(HF$_{\rm diag}$) starting from Hartre-Fock
like eigenvalues provides equivalent results. Both the GW and G$_0$W$_0$(HF$_{\rm diag}$) approaches
provide ionization energies with a mean average error within $\sim$ 0.3 eV ($\sim$ 4$\%$) of 
the experiment. Concerning the HOMO-LUMO gaps, with a  limited number of experimental data, the 
same GW and G$_0$W$_0$(HF$_{\rm diag}$) approaches yield a mean average error of  0.1-0.2 eV
(2-4$\%$), in much better agreement than the 4.1 eV (71$\%$) error within DFT/LDA,
but also in significantly better agreement than the 0.76 eV ($13\%$) error  within the 
``standard" G$_0$W$_0$(LDA) approach. 
The possibility of performing GW calculations for molecules comprizing several 
dozens of atoms with reasonnable computer time and accuracy, with a scheme allowing 
to obtain the full quasiparticle spectrum of both finite size and extended systems, 
opens the way to the investigation of organic photovoltaic systems with techniques 
that may possibly compete with well-established quantum chemical approaches.

\textbf{Acknowledgements -}
X.B. is indebted to Marc Cassida for suggesting numerous references, Pascal Quemerais 
for pointing out the relations between the radial integrals involved in the Coulomb matrix 
elements evaluation and the $_1F_1$ confluent hypergeometric functions, Julian Gale for
discussions on the gaussians fit of the radial part of numerical orbitals, and Brice Arnaud
for suggesting techniques to stabilize the imaginary axis integration in the contour
deformation approach. 
Calculations have been performed on the CIMENT plateform in Grenoble thanks to the Nanostar RTRA project.



\begin{thebibliography}{150bliography}
\bibitem{choice}
The availability of experimental results for organic molecules in the gas phase is also
an important factor directing our choice.
\bibitem{TZDP}
In good agreement with the results of Ref.~\onlinecite{Rostgaard10} in the case of
small molecules, we have verified in the case of the polyacenes that a triple-$\zeta$
+ double polarization (TZDP) basis
for expending the Kohn-Sham orbitals increases the quasiparticle correction to the
ionization energy by less than 0.07 eV as compared to the DZP basis adopted in the
present study.
\bibitem{localization}
The cutoff radius used by the SIESTA package is chosen large enough so that
the ``normalization spilling" induced by truncating the gaussian tails is less 
that $10^{-10}$.
\bibitem{b3lyp}
The B3LYP/6-31G(d) HOMO-LUMO gap of C$_{60}$ was found to be 2.77 eV in 
Refs.~\onlinecite{ShucklaC60,ZhangC60} and that of the H$_2$P porphyrin was shown
to be 2.9 eV in Ref.~\onlinecite{Nguyen}.
\bibitem{convergency}
We observe that our largest discrepancies with planewave based $G_0W_0$(LDA) calculations occurs
for rather large systems. Further work is needed to know is such discrepancies stem from
the gaussian basis or possibly from the difficulty in converging planewave calculations in 
terms of unit cell and dielectric matrix size or number of conduction bands. 
Tests on benzene with the PWSCF/Yambo packages \cite{PWSCF,Yambo} show that too small a 
cutoff on the dielectric matrix size, or the summation over a too limited number of conduction
bands, open the band gap as a result both of underscreening (unconverged W) and by neglecting
correlation with higher states (unconverged G).
\bibitem{homom1}
Apart form the band gap, the opening up to 1.5 eV for the HOMO/HOMO-1 energy difference in PTCDA,
(instead of 0.7 eV LDA value), and up to 1.2 eV for the HOMO/HOMO-2 energy difference in H$_2$P 
(instead of 0.83 eV LDA value) are in excellent agreement with the planewave $G_0W_0$(LDA)
calculations of Refs.~\onlinecite{Dori06,Palumno09}.
\end{thebibliography}
\end{document}